\documentclass[useAMS,usenatbib]{mn2e}
\usepackage{amssymb}
\usepackage{graphicx}
\usepackage{xspace}

\usepackage{amssymb,latexsym,graphicx,natbib,eufrak,times,amsmath}

\usepackage{ifthen}
\def\draftversion{0} 

\makeatletter
\newcommand\mytoc{%
    \@starttoc{toc}%
}
\makeatother

\ifthenelse{\equal{\draftversion}{0}}{
	\usepackage{xcolor}
	\newcommand{\tmp}{}
	\newenvironment{envcomm}[1]{\renewcommand{\tmp}{#1}\begin{color}{blue}\begin{center}\hrule\vspace{0.5mm}\tmp's COMMENTS\end{center}}{\begin{center}END OF \tmp's COMMENTS\vspace{0.5mm}\hrule\end{center}\end{color}}
	\newenvironment{draft}{\begin{color}[rgb]{0,0.4,0}\begin{center}\hrule\vspace{0.5mm}DRAFT\end{center}}{\begin{center}END OF DRAFT\vspace{0.5mm}\hrule\end{center}\end{color}}
	\newcommand{\comcomm}[2]{\begin{color}{blue}\ $\bullet$ \textbf{#1:} #2 $\bullet$\ \end{color}}
	\newcommand{\revend}[1]{\par\begin{color}[rgb]{0,0.4,0}\begin{center}\hrule\vspace{0.5mm}END OF #1's REVISIONS\vspace{0.5mm}\hrule\end{center}\end{color}\par}
	\newcommand{\todo}[1]{\begin{color}{red}\ $\bullet$ \textbf{To do: }#1 $\bullet$\ \end{color}}
	
	\newcommand{\del}[1]{\begin{color}[rgb]{0,0.5,0.0}\ $\bullet$ \textbf{Deleted: }#1 $\bullet$\ \end{color}}
	\newcommand{\sk}[1]{\begin{color}[rgb]{0.6,0,0.6}#1\end{color}}
	\newcommand{\toc}{\par\begin{color}[rgb]{0.6,0,0.6}\begin{center}\hrule\vspace{0.5mm}\begingroup\small\let\cleardoublepage\relax\let\clearpage\relax\mytoc\endgroup\vspace{0.5mm}\hrule\end{center}\end{color}\par}
	}{
	\newsavebox{\trashcan}
	\newenvironment{envcomm}[1]{\begin{lrbox}{\trashcan}\begin{minipage}{\columnwidth}}{\end{minipage}\end{lrbox}}
	
	\newcommand{\comcomm}[2]{}
	\newcommand{\revend}[1]{}
	\newcommand{\todo}[1]{}
	
	\newcommand{\del}[1]{}
	\newcommand{\sk}[1]{}
	\newcommand{\toc}{}
	}

\long\def\symbolfootnote[#1]#2{\begingroup%
\def\thefootnote{\fnsymbol{footnote}}\footnote[#1]{#2}\endgroup} 

\newcommand{\apj}{ApJ}
\newcommand{\aap}{A\&A}
\newcommand{\mnras}{MNRAS}
\newcommand{\pasj}{PASJ}

\newcommand{\mh}{\ensuremath{\textrm{\,--\,}}}
\newcommand{\bb}[1]{\ifmmode \mbox{\boldmath $ #1$} \else  \mbox{\boldmath $#1$} \fi}

\newcommand{\dd}{\ensuremath{\,\mathrm{d}}}

\newcommand{\U}[1]{\ensuremath{\mathrm{~#1}}}

\newcommand{\Myr}{\U{Myr}}

\newcommand{\pc}{\U{pc}}
\newcommand{\kpc}{\U{kpc}}

\newcommand{\msun}{\U{M}_{\odot}}
\newcommand{\Msun}{\msun}

\newcommand{\kms}{\U{km\ s^{-1}}}

\newcommand{\tidaltensor}{\textbf{T}}

\newcommand{\nbtt}{\texttt{NBODY6tt}\xspace}

\newcommand{\exi}{\textbf{\emph{e}}_i}
\newcommand{\exj}{\textbf{\emph{e}}_j}

\newcommand{\nbody}{\texttt{NBODY6}\xspace}

\newcommand{\order}{\mathcal{O}}

\newcommand{\eqn}[2][]{Equation#1~\ref{eqn:#2}} 
\newcommand{\fig}[2][]{Figure#1~\ref{fig:#2}}

\newcommand{\sect}[2][]{Section#1~\ref{sec:#2}}
\renewcommand{\eqn}[2][]{equation#1~(\ref{eqn:#2})}
\renewcommand{\fig}[2][]{Fig#1.~\ref{fig:#2}}

\title[NBODY6tt]{A flexible method to evolve collisional systems and their tidal debris in external potentials}

\author[Renaud \& Gieles] {Florent~Renaud$^{1,2}$\thanks{f.renaud@surrey.ac.uk} and Mark~Gieles$^1$\\
$^1$ Department of Physics, University of Surrey, Guildford, GU2 7XH, UK\\
$^2$ Laboratoire AIM Paris-Saclay, CEA/IRFU/SAp, Universit\'e Paris Diderot, F-91191 Gif-sur-Yvette Cedex, France
}

\date{Accepted 2015 February 4. Received 2015 January 28; in original form 2014 December 22}

\begin{document}
\maketitle


\begin{abstract}
We introduce a numerical method to integrate tidal effects on collisional systems, using any definition of the external potential as a function of space and time. Rather than using a linearisation of the tidal field, this new method follows a differential technique to numerically evaluate the tidal acceleration and its time derivative. Theses are then used to integrate the motions of the components of the collisional systems, like stars in star clusters, using a predictor-corrector scheme. The versatility of this approach allows the study of star clusters, including their tidal tails, in complex, multi-components, time-evolving external potentials. The method is implemented in the code \nbody \citep{Aarseth2003}.
\end{abstract}

\section{Introduction}

Recent developments in algorithms and hardware open new perspectives in the treatment of the collisional $N$-body problem. For example, it is now possible to model the long-term evolution of massive globular clusters with up to several $10^5$ stars \citep{Heggie2011, Heggie2014, Hurley2012}. Combined with much faster methods allowing a wide exploration of the parameter space (e.g. \citealt{Joshi2000}, \citealt{Alexander2012}, \citealt{Hypki2013}, \citealt{Sollima2014}, \citealt{Vasiliev2015}), $N$-body simulations are reaching a high degree of realism, and can even be used to numerically reproduce real globular clusters \citep{Heggie2014}.

The old age of globular clusters usually requires to run simulations covering an entire Hubble time to probe their full evolution. However, cosmological simulations tell us that, during such a period of time, the galactic environments of these clusters undergo secular (accretion) and violent (galaxy interactions and mergers) evolutions \citep[among many others, see e.g.][]{Agertz2011}. The complexity of these external potentials is often neglected in $N$-body simulations of clusters. However, several studies have adressed this problem, either by arbitrarily switching tidal effects to mimic the accretion of a dwarf satellite onto a massive galaxy \citep{Miholics2014, Bianchini2015}, or by (partly) coupling galaxy simulations to star cluster simulations \citep{Fujii2007, Renaud2013a, Rieder2013}.

Among other approaches, in \citet{Renaud2011}, we proposed a method to extract the tidal information (in the form of tables of tensors) along an orbit in a galaxy or cosmology simulation. This method allows any kind of galactic potential (and cluster orbit) to be considered, including complex time-dependent ones like those found in galaxy mergers (see an application in \citealt{Renaud2013a}). It however suffers from three main limitations.
\begin{enumerate}
\item The need to run a galaxy simulation first. This can be numerically costly, time consuming and thus crippling for some users. Furthermore, running a full galaxy simulation is unnecessary when the evolution of the galaxy can be described analytically. (For example, the secular mass growth of a galaxy can be mimicked, at first order, by scaling the total mass and scale radii of the galaxy without modifying the shape of its potential, see \citealt{Diemer2013, Buist2014}.)
\item The different timescales between galactic and cluster scales. The tidal information is generally sampled at a much coarser frequency ($\sim 1\mh 5 \Myr$ and $\sim 10\mh 50 \pc$) than what is required in cluster simulations. Therefore the evaluation of this information at the time and position where it is needed by the cluster simulation requires interpolations of the data available.
\item The so-called tidal approximation, i.e. the linearisation of the tidal forces. This introduces errors at large distances from the cluster centre and thus forbids the study of tidal tails.
\end{enumerate}

In this paper, we propose an alternative version which overcomes these limitations. In this new method, the user provides a program routine returning the galactic potential as a function of position and time. (Running a galaxy simulation beforehand is \emph{not} required, and the information on the external potential is not discretised, which circumvents the need for interpolations.) The code uses this routine to compute the motion of the cluster, the tidal acceleration on its stars and the relevant derivatives (used to increase the accuracy, as described in \sect{method}). Since the galactic potential is known at all possible positions, the tidal acceleration can be added to the motion of every star in the simulation, including those in tidal debris, which eliminates the limitation of the tidal approximation.

In the present version, the external potential must be defined by the user, in a form of a numerical code routine. It can be an simple analytical function of position and time, and/or involve the numerical solving of more complex functional forms. This allows for a wide diversity of cases, from spherically symmetric static models to time-evolving, triaxial, multi-component models. In some cases however, describing the potential with a function is too involved (e.g. galaxy mergers), and it would be preferable to follow the approach of \citet{Renaud2011}, i.e. tables of tensor coefficients. Both methods have been implemented in \nbody and its GPU version, \citep{Aarseth2003, Nitadori2012}, under the name \nbtt, and are available online\footnote{{\tt http://personal.ph.surrey.ac.uk/$\sim$fr0005/nbody6tt.php}}.

\section{Method}
\label{sec:method}

In the rest of the paper, we describe the new method in the context of modelling a star cluster in a galactic potential. The method can however be used in other configurations, when an external potential must be included in a collisional $N$-body system.

\subsection{Direct or differential?}
\label{sec:diff}

The evaluation of the contribution of the galaxy on a star cluster can be done in two ways.
\begin{enumerate}
\item Direct approach: the galactic acceleration on a star is added to that from the $N-1$ other stars of the cluster. In that case, the coordinate frame is centered on the galaxy.
\item Differential approach: the contribution of the galaxy is evaluated at the position of the star and at the position the cluster (usually its center of mass), and the difference is computed. The motion of the star is thus integrated with respect to the cluster, which has its own motion around the galaxy. The differential terms are called ``tidal''.
\end{enumerate}

Because the galactic contribution on a star (e.g. on its acceleration) can be several orders of magnitude different than that from the cluster, the first method which consists in summing the two contributions can lead to numerical errors. For this reason, we adopt the second approach (as in all versions of \nbody and \nbtt).

\subsection{Numerical derivatives}
\label{sec:derivatives}

When the galactic potential yields an analytical description, the most accurate approach to include tidal effects would be to derive the galactic acceleration (and higher order terms) analytically. In some cases however, these derivations could be rather involved and even dissuasive. Such situations are encountered for complex functional forms of the potential (asymmetric, time-dependent etc), or in the rarer cases when the potential (or its derivatives) requires a numerical evaluation. If accurate enough, a numerical approach to compute both the orbit of the cluster and the tidal acceleration on its stars is often more appropriate and/or convenient. In the following Sections, we propose such a method and evaluate the error introduced. We found that, when solving the $N$-body problem on single precision hardware (e.g. on Graphical Processing Unit, GPU) as it is often the case, the error introduced by our method gets truncated, such that our approach is as accurate as an analytical derivation, at least for the cases considered below (see \sect{tests}).

\begin{figure}
\includegraphics{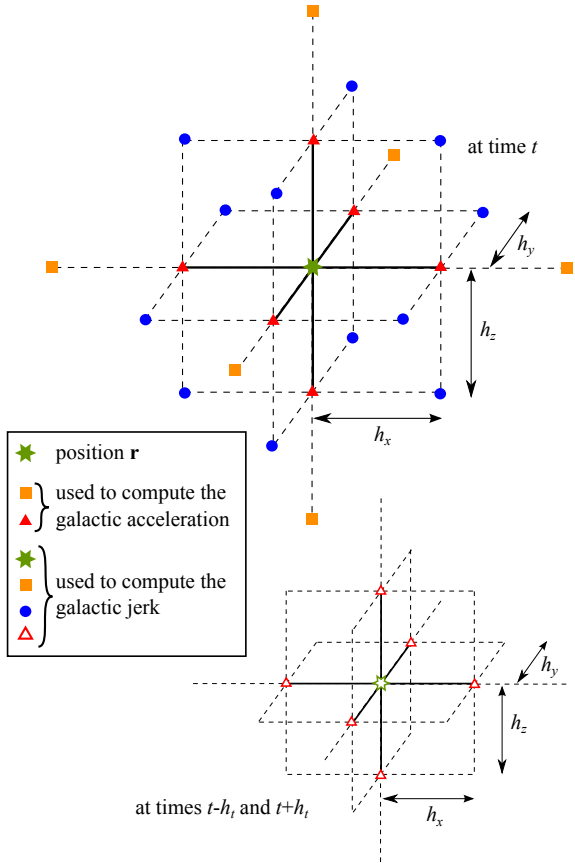}
\caption{Stencil used to compute the first (acceleration) and second (jerk) order derivatives of the galactic potential at a given position (green star).}
\label{fig:scheme}
\end{figure}

In our method, a routine returns the galactic potential $\phi_\textrm{G}$ as a function of position $\bb{r}$ and time $t$. As in the 2011 version of \nbtt, all computations are done in an inertial reference frame, where there are no fictitious forces (centrifugal, Coriolis and Euler, which are \emph{a priori} unknown in the general case). The acceleration (per unit of mass) from the galaxy at the position $\bb{r} = \sum x_i \exi$ and time $t$ is thus given by minus the gradient of the potential, i.e.
\begin{equation}
\bb{a_\textrm{G}}(\bb{r}, t) = -\frac{\dd \phi_\textrm{G}(\bb{r}, t)}{\dd \bb{r}}.
\end{equation}
Numerically, we evaluate its $i$-th component using a central finite difference (see \fig{scheme}) with a fourth-order accuracy:
\begin{align}
a^i_\textrm{G}(\bb{r}, t) \approx& - (12h_i)^{-1} \left[\phi_\textrm{G}(\bb{r}-2h_i\exi, t) - 8\phi_\textrm{G}(\bb{r}-h_i\exi, t) \right. \nonumber\\
                         & \left. + 8\phi_\textrm{G}(\bb{r}+h_i\exi, t) - \phi_\textrm{G}(\bb{r}+2h_i\exi, t) \right],
\label{eqn:acceleration}
\end{align}
which thus has errors $\order(h_i^4)$. The choice of the value of the step size $h_i$ is described in \sect{ttdx}.

The time derivative of the galactic acceleration, or jerk ($\bb{j_\textrm{G}}$), is then computed using the relation
\begin{align}
\bb{j_\textrm{G}}(\bb{r}, t) = \frac{\dd \bb{a_\textrm{G}}(\bb{r}, t)}{\dd t} & = \frac{\partial \bb{a_\textrm{G}}(\bb{r}, t)}{\partial \bb{r}} \frac{\partial \bb{r}}{\partial t} + \frac{\partial \bb{a_\textrm{G}}(\bb{r}, t)}{\partial t} \nonumber \\
	& = \tidaltensor(\bb{r}, t) \bb{v} + \frac{\partial \bb{a_\textrm{G}}(\bb{r}, t)}{\partial t},
\label{eqn:jerk}
\end{align}
where $\bb{v}$ is the velocity vector of the star with respect to the galaxy. The space derivative of the galactic acceleration (i.e. minus the Hessian matrix of the galactic potential) is the tidal tensor $\tidaltensor$, which we compute numerically by using a second order central finite difference of $\phi_\textrm{G}$, with a second-order accuracy (\fig{scheme}). Its components read
\begin{align}
T^{ij}(\bb{r}, t) \approx - (4h_ih_j)^{-1} \left[ \right. & \phi_\textrm{G}(\bb{r}-h_i\exi+h_j\exj, t) \nonumber \\
	- & \phi_\textrm{G}(\bb{r} +h_i\exi + h_j\exj, t) \nonumber \\
	+ & \phi_\textrm{G}(\bb{r} + h_i\exi -h_j\exj, t) \nonumber \\
	- & \phi_\textrm{G}(\bb{r} -h_i\exi -h_j\exj, t) \left. \right].
\label{eqn:tensor}
\end{align}
The partial time derivative of the acceleration, computed with a first order finite difference of $\bb{a_\textrm{G}}$ with a second-order accuracy, is
\begin{align}
\frac{\partial \bb{a_\textrm{G}}(\bb{r}, t)}{\partial t} \approx & (2h_t)^{-1} \left[ - \bb{a_\textrm{G}}(\bb{r}, t-h_t) + \bb{a_\textrm{G}}(\bb{r}, t+h_t) \right]
\label{eqn:jerktime}
\end{align}
where $h_t$ is the step size for the time dimension. Here, the fourth-order accuracy provided by \eqn{acceleration} is unnecessary. Instead, we compute the acceleration with a second-order accuracy:
\begin{align}
a^i_\textrm{G}(\bb{r}, t) \approx& - (2h_i)^{-1} \left[-\phi_\textrm{G}(\bb{r}-h_i\exi, t) + \phi_\textrm{G}(\bb{r}+h_i\exi, t) \right],
\end{align}
such that the $i$-th component of its partial time derivative reads
\begin{align}
\frac{\partial a^i_\textrm{G}(\bb{r}, t)}{\partial t} \approx -(4h_ih_t)^{-1} \left[ \right. &\phi_\textrm{G}(\bb{r}-h_i\exi, t-h_t) \nonumber \\
- &\phi_\textrm{G}(\bb{r}+h_i\exi, t-h_t) \nonumber \\
- &\phi_\textrm{G}(\bb{r}-h_i\exi, t+h_t) \nonumber \\
+ & \phi_\textrm{G}(\bb{r}+h_i\exi, t+h_t) \left. \right].
\label{eqn:acceleration_time}
\end{align}
For simplicity, we use the same values of the spacial step sizes $\{h_i\}$ at $t$, $t-h_t$ and $t+h_t$. Finally, the jerk is evaluated by replacing equations~(\ref{eqn:tensor}) and (\ref{eqn:acceleration_time}) in equation~(\ref{eqn:jerk}).

\subsection{Cluster orbit}

The motion of the cluster around the galaxy is described using a guiding centre, as already done in \nbody by setting the option KZ(14) to 3 or 4. This pseudo-particle initially matches the centre of mass of the cluster, but can slightly deviate from it later on, as stars tidally ejected from the cluster take away momentum in an asymmetric fashion. \nbtt integrates the equation of motion of the guiding center using the galactic acceleration (equation~\ref{eqn:acceleration}) and jerk (equation~\ref{eqn:jerk}) in a predictor-corrector scheme \citep{Aarseth2003}.

Note that dynamical friction is not included in the integration of the cluster orbit (but see Petts et al., in preparation).

\subsection{Tidal acceleration}

When the regular force on a star (either in the cluster or in the tidal debris) must be updated, the contribution of the galaxy is evaluated, using the differential approach presented in \sect{diff}. The galactic acceleration (resp. jerk) at the position of the guiding center of the cluster is subtracted from that at the position of the star. We thus obtain the tidal (i.e. differential) acceleration (resp. jerk), used in the predictor-corrector integration scheme of \nbody to evolve the motion of the star in the cluster. The advantage of this approach is that it allows us to follow the motion of the stars on galactic orbits after they have left the cluster.

\subsection{Step sizes}
\label{sec:ttdx}

In this Section, for simplicity, we consider the derivation of the potential $\phi_\textrm{G}$ with respect to a single dimension $x$, with the step size $h$, and we will generalize our approach to the multi-dimensional case later.

\begin{figure}
\includegraphics{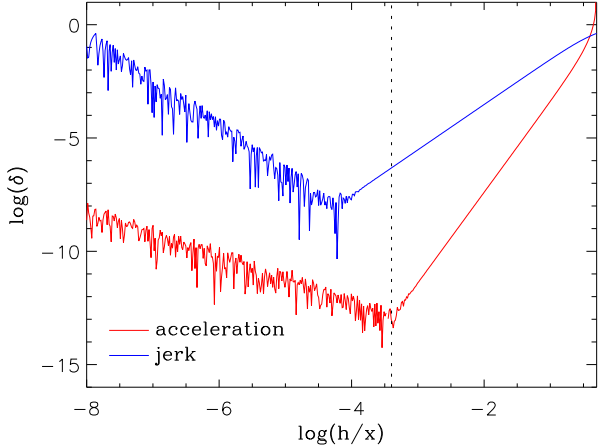}
\caption{Relative numerical error (1 - numerical/exact) made on the acceleration (red) and the jerk (blue) from a point-mass potential, as a function of the step size (for $x = 1$). The vertical dotted line marks the choice of $h$ adopted in our implementation (equation~\ref{eqn:h}).}
\label{fig:ttdx}
\end{figure}

The accuracy in the evaluations of the numerical derivatives presented in \sect{derivatives} rely on the choice of step sizes. A too small value would lead to round-off errors while the derivative would not be accurate for a too large value. According to \citet[their Section 5.7]{Press2007}, the optimum value of $h$ depends on the ``curvature scale'' $x_c$ of the potential at the position it is evaluated, as
\begin{equation}
h \sim \epsilon^\zeta x_c,
\label{eqn:stepsize}
\end{equation}
where $\epsilon$ is comparable to the machine accuracy (i.e. $\sim 10^{-16}$ in double precision). Both $\zeta$ and $x_c$ depend on the order of the derivative considered. By writing the Taylor series expansion of $\phi_\textrm{G}(x+h)$ and by seeking the value of $h$ which minimizes the sum of the round-off and truncation errors (see an example at lower order in \citealt{Press2007}, their Equation 5.7.5), one can show that 
\begin{equation}
\zeta = \frac{1}{5} \qquad \textrm{and} \qquad x_c = \left(\frac{\phi_\textrm{G}}{\frac{\partial^5 \phi_\textrm{G}}{\partial x^5}}\right)^{1/5}
\end{equation}
for equation~(\ref{eqn:acceleration}), and 
\begin{equation}
\zeta = \frac{1}{3} \qquad \textrm{and} \qquad x_c = \left(\frac{\phi_\textrm{G}}{\frac{\partial^3 \phi_\textrm{G}}{\partial x^3}}\right)^{1/3}
\end{equation}
for equations~(\ref{eqn:tensor}) and (\ref{eqn:acceleration_time}). Therefore, computing the optimum step size requires, paradoxically, the evaluation of higher order derivatives of the potential. In principle, this could be done through the approach of \citet{Ridders1982}, at the cost of many additional evaluations of the potential. Once the optimum step size found, we could build the stencil of \fig{scheme} and compute all the relevant derivatives, for all stars in the cluster. In practice, this methods appears to be of limited interest with respect to its significant extra numerical cost. Following the advice of \citet{Press2007}, we choose to adopt a much simpler and faster approach and assume that the curvature scale can be approximated by the position $x$. In that case, \fig{ttdx} shows the relative numerical error made on the acceleration and the jerk using the potential of a point-mass ($\phi_\textrm{G} = -1/x$). To take avantage of the stencil method (\fig{scheme}), and to further limit the number of evaluations of the potential, we choose to use the same step size for both the computation of the acceleration and the jerk. Since the jerk is only used in the predictor corrector method, the accuracy on its value is less critical than that of the acceleration\footnote{The jerk is multiplied by the small time step of the predictor-corrector scheme before being added to the acceleration. Therefore, it is the error on the jerk times this time step (typically $10^{-6}$ the orbital period) which must be compared to the error on the acceleration, such that the dominant source of errors is that made on the acceleration.}. Therefore, we have adopted the optimum step size for the acceleration to compute both the acceleration and the jerk.

Setting $h \propto x$ implies numerical issues for $x=0$ leading to infinite derivatives. To limit the number of occurrences of this situation, we choose instead to use $h \propto r$. The step size would then be far from being optimum where, e.g. $x \ll r$, but the error would generally be made on a small component of the total acceleration ($a^x_\textrm{G} \ll ||\bb{a}_\textrm{G}||$). In the end, we adopt:
\begin{equation}
h_i = 4 \times 10^{-4} r
\label{eqn:h}
\end{equation}
for all three values of $i$, and use this empirical relation for all potentials. We note however that our choice of the step size would not be optimum at the vicinity of substructures in the potential, like spirals arm in a galactic disc. 

For time-dependent potentials, the time step $h_t$ is taken to be the same as the time step of the predictor corrector scheme used when integrating the motion of the guiding centre in the galaxy. It is the same for all stars.

This method is tested in \sect{tests}.

\subsection{Numerical precision}
\label{sec:precision}

In the implementation of our method in \nbody, the computations of the external potential and its derivatives are performed in double-precision to minimize the impact of the loss of accuracy during the numerical derivations.

We note however that the actual force might be truncated to single-precision in some cases. The majority of the GPUs used in the community are limited to single-precision and thus introduce an ``hardware truncation'' of the accuracy of the numerical variables they manipulate. Such truncation affects the evaluation of the regular force in \nbody, to which the tidal force is added. Therefore, despite being evaluated in double-precision, the tidal force applied to the particles are truncated to single-precision when using GPUs. This limitation does not concern simulations run on CPUs.

\subsection{Energy conservation}

One way of monitoring the numerical errors made during a simulation is to control the conservation of energy. However, time-dependent tides imply that the energy is not conserved. The code circumvented this issue through the following method\footnote{already used in previous versions of \nbody when setting the option \texttt{KZ(14)=3}.}. Let $\dot{W}$ be the time derivative of the internal energy of the cluster (potential energy $U$ from internal interactions plus kinetic energy $K$ relative to the guiding centre of the cluster). It can be written as
\begin{align}
\dot{W} \equiv \frac{\dd (U+K)}{\dd t} =& \sum_i^N \frac{\dd U}{\dd \bb{r}_i} \frac{\dd \bb{r}_i}{\dd t} + \sum_i^N m_i \bb{v}_i \frac{\dd \bb{v}_i}{\dd t} \nonumber\\
=& \sum_i^N - m_i \bb{a_\textrm{C}}_i \bb{v}_i + \sum_i^N m_i \bb{v}_i (\bb{a_\textrm{C}}_i + \bb{a_\textrm{G}}_i) \nonumber \\
=& \sum_i^N m_i \bb{v}_i \bb{a_\textrm{G}}_i,
\label{eqn:wdot}
\end{align}
where $\bb{a_\textrm{C}}$ represents the internal acceleration due to the stars in the cluster, $m_i$ is the mass of the $i$-th star, and the sums are made over all stars in the system. We note that the second time derivative of $W$ reads
\begin{equation}
\ddot{W} = \sum_i^N m_i \left[ (\bb{a_\textrm{C}}_i + \bb{a_\textrm{G}}_i) \bb{a_\textrm{G}}_i + \bb{v}_i \bb{j_\textrm{G}}_i \right].
\label{eqn:w2dot}
\end{equation}
The definition of $\dot{W}$ implies that
\begin{equation}
U+K- \int \dot{W} \dd t = \textrm{constant},
\label{eqn:cste}
\end{equation}
which is the assertion that the code must verify.

In practice, to numerically obtain the variation ($\dot{W}$) of internal energy between the timesteps $t_1$ and $t_0$, we compute the Taylor series of $W$ at the time $t_0$, truncated to the second order:
\begin{equation}
W(t_1) \approx W(t_0) + \frac{\dd W}{\dd t} (t_1-t_0) + \frac{\dd^2 W}{\dd t^2} \frac{(t_1-t_0)^2}{2},
\end{equation}
whence
\begin{equation}
\Delta W \approx \dot{W} \Delta t + \ddot{W} \frac{(\Delta t)^2}{2}.
\end{equation}
Using equations~(\ref{eqn:wdot}) and (\ref{eqn:w2dot}), we can compute the variation $\Delta W$ at a given timestep in the simulation. The accumulation of these variations since the beginning of the simulation gives the numerical equivalent of $\int \dot{W} \dd t$ that we subtract to the value of $U+K$ to check \eqn{cste}\footnote{From \eqn{cste}, one can see that $-\int \dot{W} \dd t = -W$ is the part of the tidal energy that contributes to energy conservation. It is \emph{not} the full tidal energy in the general case.}.

\section{Tests}
\label{sec:tests}

In this Section, we compare the results from our implementation of the method to either analytical solutions or numerical results from \nbody, and thus do not explore the full possibilities offered by the new method.

\subsection{Acceleration}
\label{sec:acceleration}

\begin{figure}
\includegraphics{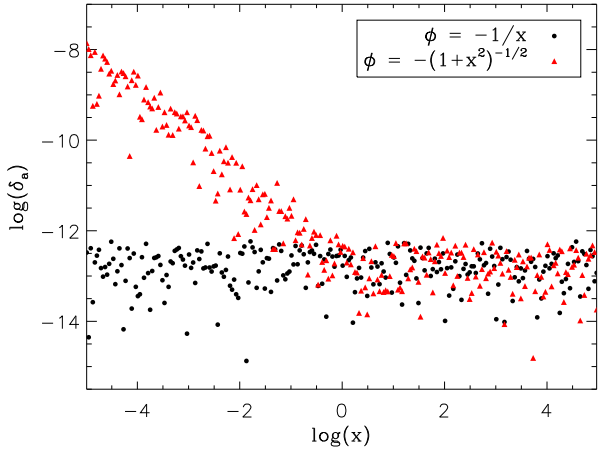}
\caption{Relative numerical error (1 - numerical/exact) on the $x$-component of the acceleration, as a function of $x$. The potentials adopted have the form $-1/x$ (black) and $-(1+x^2)^{-1/2}$ (red). The error remains small over the range considered, despite an increase when the potential becomes flatter and flatter.}
\label{fig:acceleration}
\end{figure}

\fig{acceleration} shows the relative numerical error made on the $x$-component of the acceleration (evaluated with equation~\ref{eqn:acceleration}). We monitor the accuracy of the derivation scheme by both setting divergent (cusped) and cored potentials, such that a wide variety of potential slopes are considered. The relative error on the acceleration computed in double-precision overcomes $10^{-7}$, i.e. the machine accuracy in single-precision, when it is computed at a distance to the centre of the potential $\sim 3\times10^6$ times shorter than the potential characteristic scale. In the context of cored potentials of galaxies, this would corresponds to cluster-galaxy distances of a few parsecs, i.e. a relatively rare situation.

Therefore, except in the case previously mentioned, following the discussion in \sect{precision}, the net accuracy of the acceleration would be set by the machine precision when using single-precision GPUs. In that view, the acceleration evaluated with our method could be considered as accurate as a computation using the analytical expression.

\subsection{Cluster orbit}
\label{sec:orbit}

\begin{figure}
\includegraphics{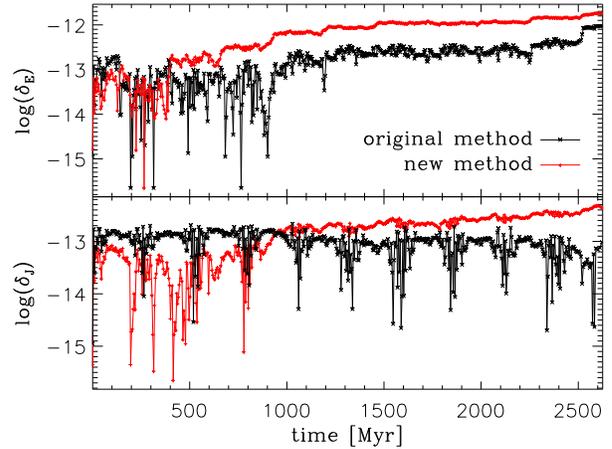}
\caption{Relative numerical errors (1 - numerical/exact) on the orbital energy (top) and angular momentum (bottom), of the guiding centre of a cluster on an eccentric orbit (eccentricity 0.5, apocenter = $3 \kpc$) around a point-mass galaxy ($10^9 \Msun$).}
\label{fig:energy}
\end{figure}

\fig{energy} shows the numerical errors made in the energy and angular momentum of the guiding center of a cluster on an eccentric orbit of eccentricity 0.5 and apocenter distance $3 \kpc$ around a point mass galaxy ($10^9 \Msun$). This corresponds to an orbital period of $\sim 260 \Myr$. The errors from the new method remain of the order of those from the original method. A long-term drift exists but is limited to about one dex over 10 orbital periods, indicating that both quantities are well conserved over periods of time matching the typical life-time of clusters, or the typical duration of these simulations.

\subsection{Cluster evolution}

\begin{figure}
\includegraphics{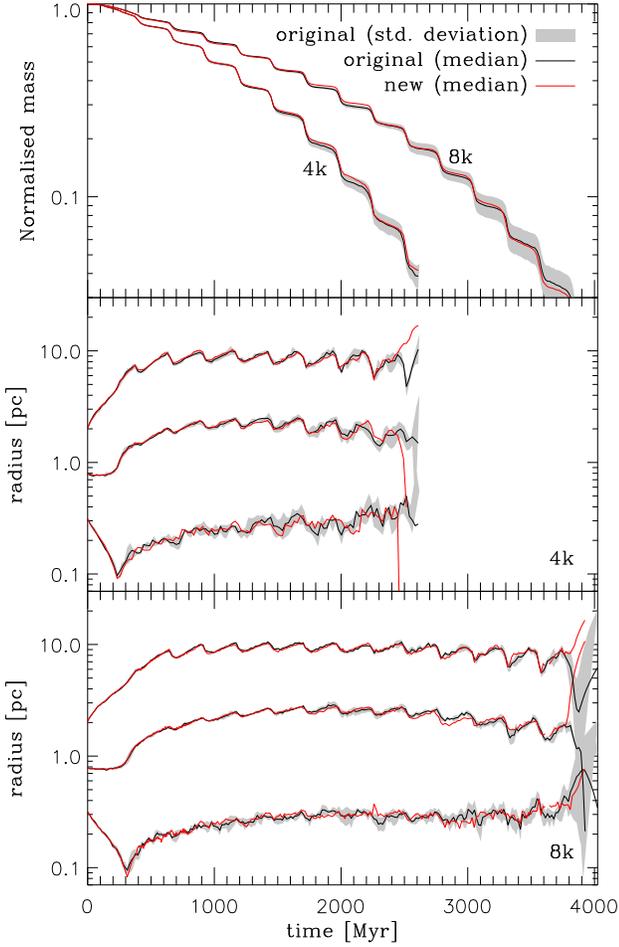}
\caption{Evolution of the mass and the structure (shown with the 10\%, 50\% and 90\% Lagrange radii) of models of a clusters with initially 4096 and 8192 stars (labelled 4k and 8k respectively) integrated with the new method, and compared to the \nbody original method. Each curve represents the median value over four $N$-body realisations of the initial conditions, and the shaded areas show the standard deviations of these realisations (original method only), following \citet{Ernst2011} and \citet{Whitehead2013}. Strong variations of the radii near the time of dissolution of the clusters are due to uncertainties in the determination of the centre of small-$N$ systems.}
\label{fig:cluster}
\end{figure}

\fig{cluster} compares the evolution of the mass and a few Lagrange radii of clusters modelled with the new method with that of the same clusters modelled with \nbody. The clusters are equal-mass models of 4096 or 8192 stars ($1 \Msun$ each) on a Plummer profile with an initial virial radius of $1 \pc$, and placed on an eccentric orbit around a point-mass galaxy, as described in the previous Section. The quantities plotted are computed using the bound stars, i.e. those for which the sum of the kinetic and internal energy (with respect to the guiding centre) is negative, as in \citet{Renaud2011}. (In other words, we neglect the tidal energy when deciding the membership of stars.) To evaluate the amplitude of Poisson's noise in our measurements, we have run simulations of four realisations of each cluster by changing the seed of the random number generator used to produce the initial conditions. 

The evolution of the mass and size of the clusters is compared to that computed using the built-in version of \nbody (option \texttt{KZ(14)=3}). We note that the low-frequency evolution of the clusters, mainly connected to the orbital period, is well reproduced by the new implementation until the end of the simulations. As expected, the largest deviations between the two methods are found in the inner regions of the cluster, where the evolution is the most vulnerable to statistical effects leading to the stochastic formation and destruction of binaries. The amplitude of the deviations increases as the number of stars shrinks, as a result of Poisson noise.

In conclusion, the agreement between the two methods is very good for the entire lifetime of the clusters considered, and over all phases of their orbit (apocenter, pericenter and in between).

\subsection{Steep potentials and tidal shocks}

To further test the method, we consider the case of a cluster plunging through a disc. The external potential is modelled with a single \citet{Miyamoto1975} disc:
\begin{equation}
\phi_\textrm{G} = -\frac{GM}{\sqrt{x^2 + y^2 + \left(a+\sqrt{z^2+b^2}\right)^2}},
\label{eqn:mn}
\end{equation}
with the parameters $M = 10^{11} \msun$, $a = 5 \kpc$ and $b = 300 \pc$. The cluster is initially set at the position $x_0 = y_0 = 0$, $z_0 = 1.5 \kpc$, i.e. above the median plane of the disc, with the initial velocity toward the disc $v_{z,0}= -150 \kms$. Such configuration can be setup in the original version of \nbody (option {\tt KZ(14)=3}) and thus we can compare here again the result from the new method to that of \nbody. The cluster is made of 8196 equal-mass stars distributed on a Plummer profile with an initial virial radius of $3 \pc$. The cluster crosses the median plan of the disc with a velocity of $\approx 230 \kms$ and takes $\approx 2.6 \Myr$ to cover the $2\times 300 \pc$ of the disc scale-height.

\begin{figure}
\includegraphics{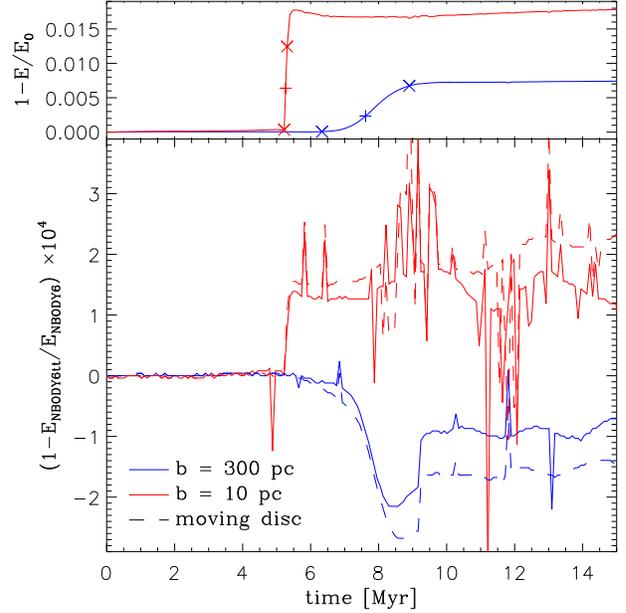}
\caption{Top: evolution of the internal energy (potential + kinetic), normalised to its initial value, of a cluster plunging through a \citet{Miyamoto1975} disc of scale-heigh $b = 300\pc$ (blue) and of $b= 10\pc$ (red). The instants when the cluster reaches the scale-heigh of the disc ($z = \pm b$) and the median plan ($z=0$) are marked with crosses and a plus-sign. Bottom: relative difference in the internal energy between the two methods. The dashed curves correspond to the setup where the initial velocity is given to the disc, and not the cluster (Section~\ref{sec:timedependentpot}). (Spikes in the curves corresponds to high velocity stars being ejected from the cluster. Such events are subject to Poison noise and thus varies from method to method.)}
\label{fig:disk_energy}
\end{figure}

\fig{disk_energy} shows the tidal heating of the cluster, and the relative difference in the internal energy of the cluster (potential + kinetic) between the two methods. The encounter leads to an increase of about one percent of the cluster initial internal energy over a few Myr. During this period, the rapidly varying external potential induces differences results from two methods. It is likely that our choice of assuming a universal curvature scale (see \sect{ttdx}) is responsible for most of the differences. However, the relative difference remains small ($\sim 10^{-4}$).

To push the code to its limits, we consider the extreme case of a cluster moving through a very thin disc. We used the same potential form as before, but with a scale-height $b = 10 \pc$ (i.e. about 3 times the initial virial radius of the cluster). We set the cluster at $z_0 = 1 \kpc$ with the same initial velocity as before ($v_{z,0} = -150 \kms$). The ``impact'' velocity is $\approx 225 \kms$, meaning that the cluster takes $\approx 0.09 \Myr$ to cover the $2\times 10 \pc$ of the disc scale-height, i.e. shorter than the cluster crossing time ($\sim 1 \Myr$). Following \citet{Spitzer1987} and \citet{Gnedin1997}, we can consider such encounter as an impulsive tidal shock, as opposed to an adiabatic effect. The comparison between the original and new methods is showed in \fig{disk_energy}. Despite an energy gain about twice larger than in the thicker disc case, the relative difference between the two method remains of the same amplitude as before.

To conclude, the difference in internal energy of a cluster computed with the two methods remains well below one percent, even in steep potentials (i.e. with a curvature scale being a strong function of position).

\subsection{Time-dependent potential}
\label{sec:timedependentpot}

Finally, we test the method in the context of a time-dependent potential. Note that, in our approach, time-dependence only affects the predictor-corrector scheme (thought the time derivative in the expression of the jerk, equation~\ref{eqn:acceleration_time}) and not the (tidal) acceleration itself which is computed at each timestep in a static way, using the present-time expression of the potential. 

\nbody does not allows for a treatment of tides with an explicit time dependence. Therefore, to compare our method to the original code, we adopt the following approach.

In the original \nbody, we use the same setup as in the previous Section, i.e. a cluster plunging through a static disc with an initial velocity $v_{z,0}$. In the new version (\nbtt), we setup the cluster with no initial velocity, and define the external potential as the same disc but moving toward the cluster with the velocity $-v_{z,0}$ by replacing $z$ with $z-v_{z,0}t$ in equation~(\ref{eqn:mn}), such that our potential becomes time-dependent. The two setups are equivalent and the physical evolution of the system should be exactly the same in both cases. The relative difference in internal energy (for the two values of $b$ adopted before, i.e. $300$ and $10 \pc$) is showed in \fig{disk_energy}. The differences induced by the disc-cluster encounter noted in the previous Section are still present here. Furthermore, the offset of the potential from the origin makes our choice of the step-size as a linear function of position (equation~\ref{eqn:h}) less optimum than before. As a consequence, the differences between the two methods are slightly amplified than for the case of the static potential centered on the origin. Despite a larger amplitude than the static cases explored before, the relative differences remains of the order of $10^{-4}$, showing that our method has a comparable behaviour than the original \nbody, even in such extreme cases.

\section{Conclusion}

We introduce a method to compute the tidal acceleration on an collisional $N$-body system embedded in an external potential which can be described with a function of position and time (analytical and/or numerical). The method evaluates the first and second space derivatives of the potential to obtain the tidal acceleration and the tidal tensor. The tensor allows us to estimate the tidal jerk which, with the acceleration, is used in a predictor-corrector scheme to integrate the equations of motion of the $N$-bodies. By circumventing the need of the classical tidal approximation (linearisation of the tidal force), this method can accurately integrate the motion of any body in the system, including those in tidal debris.

The orbit of the guiding center of the system within the external potential is also computed, following a comparable method. The numerical errors made on these quantities are of the order of $10^{-12}$ or smaller (for the test cases we considered), and thus the resulting net evolution of the $N$-body system is comparable to that obtained with other approaches. We have considered several setups where the evolution of clusters could be compared to that provided by existing methods, and found reasonable agreements in all cases. This suggests that the new method is able to produce simulations with accuracy standards close to that of \nbody.

This new method however allows a larger flexibility, as any external potential can be considered, providing it can be described by a numerical routine. Among the endless list of possible applications, one can imagine to describe the tidal effects of time-evolving multi-component galaxies including halo, bulge, disc(s), spiral pattern(s), bar(s), ring(s), and/or undergoing accretion of intergalactic gas and (to some extent) satellite galaxies. Alternative methods are however required when the external potential cannot be described by a numerical function, like for example in major galaxy mergers.

\section*{Acknowledgement}
We warmly thank Douglas Heggie and Sverre Aarseth for their most valuable input on this work and interesting discussions over the past few years, and the referee for a constructive repport. We acknowledge support from the European Research Council through grants ERC-StG-257720 and ERC-StG-335936 (CLUSTERS). MG acknowledges financial support from the Royal Society in the form of a University Research Fellowship and an equipment grant that was used to purchase the GPU machines that were used for the $N$-body computations. 
\bibliographystyle{mn2e}

\end{document}